# A *Chandra* Successor in the 2030s

**Authors:** C. Reynolds (UMD)[1], E. Hodges-Kluck (GSFC), E. Kara (MIT), M. Koss (Eureka), K. Madsen (GSFC), E. Miller (MIT), B. Williams (GSFC), S. Allen (Stanford), M. Bautz (MIT), K.Burdge (MIT), N. Cappelluti (U.Miami), B.Cenko (GSFC), L. Corrales (U.Michigan), A.Falcone (Penn State), A. Foord (UMBC), D. Haggard (McGill, Canada), L. Lopez (OSU), R. Mushotzky (UMD), A. Ptak (GSFC), H.Russell (Nottingham, UK), S. Safi-Harb (Manitoba, Canada), K.Stassun (Vanderbilt)

*For ExoPAG, CoPAG, and PhysPAG*

**0. Abstract**

*Chandra* established that soft X-ray spectral imaging with ~1” angular resolution is an essential tool for almost every area of astrophysics, from the early Universe to ice chemistry in protoplanetary disks. While *Chandra* continues to deliver groundbreaking science, it lacks the sensitivity to be the counterpart to JWST, Roman, Rubin, LISA, ngVLA, ALMA, IceCube, NewAthena, ELT, and other upcoming large facilities. **Here we highlight the scientific potential of a *Chandra* successor with >10x higher sensitivity and >50x survey speed**. While the scientific reach of such a mission is very broad, we showcase four high-priority science questions that require these capabilities – How did supermassive black holes form?, How does stellar and AGN feedback work in galaxies?, How does high-energy stellar activity affect exoplanet atmospheres? and, How do black holes merge? A *Chandra* successor in the 2030s would also demonstrate the X-ray mirror technology for a future flagship and sustain the US high energy astrophysics workforce until that flagship launches. We outline the science–instrument trade space by sketching three missions based on the NASA Probe concept study report for the *Advanced X-ray Imaging Satellite* (AXIS): **AXIS-Deep**, **AXIS-Classic**, and **AXIS-Fast**, highlighting a breakthrough science case for each and listing others that drive the design.

[1] Corresponding author, email : creynold@umd.edu

## 1. Science Investigation

We highlight four different science questions highlighted by Astro2020[1] that remain unanswered and demonstrate the importance of a *Chandra* successor with ~1" resolution and >10x the sensitivity. These questions, and others listed, showcase the science possible with deep fields, pointed surveys, and time-domain surveys. This is but a sample of the scientific potential of such a mission – we list other high-profile questions within each category, and refer the reader to the AXIS Community Science Book[2] for detailed descriptions of 138 science cases developed by more than 500 scientists.

***How did supermassive black holes form?*** **DEEP FIELD SCIENCE**

JWST has found evidence of massive ($M > 10^6\ M_\odot$) black holes at $z \sim 10$, just 500 million years after the Big Bang. As this is not enough time to grow a SMBH from the first generation of stellar remnants via Eddington-limited accretion, early SMBHs could either form from direct collapse of a gas cloud (heavy seeds) or super-Eddington accretion onto stellar-mass black holes (light seeds)[3]. X-rays – and perhaps the *absence* of X-rays[4] – are best suited to rule in favor of one scenario, as X-rays are primary tracers of accretion-powered black holes (unlike dust-processed IR emission) and can pierce the dusty shroud surrounding many, and likely most, growing black holes[5-6]. The number density of AGN at $L_X > 5\times10^{42}$ erg $s^{-1}$ at $z > 10$ can decisively answer this question[7]. Measuring the change in AGN number density from $z \sim 8$ to $z \sim 10$ can also determine which formation channel is dominant, as the seed models predict different evolution[8-9]. More broadly, an unbiased AGN survey at $z > 8$ will show how – or whether – black hole growth and galaxy growth are connected near cosmic dawn.

| Additional science questions addressed by deep fields | | |
|---|---|---|
| How quickly do SMBH binaries form and merge? | Connecting X-ray dual AGN and LISA rates | [9] |
| How coherent is SMBH accretion? | X-ray measurements of SMBH spin at z=0-2 | [10] |
| Do X-ray binaries help reionize the Universe? | X-ray vs star formation (PRIMA) at z=6-8 | [11] |

***How does stellar and AGN feedback work?*** **POINTED SURVEY SCIENCE**

We cannot understand how galaxies evolve without knowing how the energy released by massive stars and AGN regulates star formation. The input energy[12], whether radiative or mechanical, quickly goes into hot ($T > 10^6$ K) gas[13-17] that transports and disperses it far from the injection site. To quantify how efficiently this energy couples to the interstellar or circumgalactic medium, we must measure temperature and mass on the 10-1000 pc scales of the injection sites (supernova remnants, H II regions, galactic nuclei, radio bubbles, etc.). To determine how this hot gas affects its surroundings (e.g., dispersing molecular clouds[18], ejecting cool gas from the galaxy[19-20], or preventing cooling in galaxy cluster cores[13-14]) we must measure the structure of hot flows (including AGN-jet blown cavities) on scales of 0.1-10 kpc. The hot flows also carry metals that influence how stars form, live, and die, so to quantify how much feedback enriches the Universe over cosmic time, we must measure the metal content in the hot intracluster medium (ICM) in galaxy clusters, which provide a "fair sample" of the Universe, from $1 < z < 3$[21].

***How does high-energy stellar activity affect exoplanet atmospheres?***

JWST is revealing the composition of exoplanet atmospheres, setting the scene for the Habitable Worlds Observatory. Mysteries abound, including bare rock planets that appear to have lost their atmosphere for unknown reasons. Sensitive X-ray studies of Galactic open star clusters (with characteristic X-ray source separations of a few arcsec)[22] will characterize high-energy stellar activity (e.g. flares and coronal mass ejections) across spectral types as a function of age and metallicity, revealing how activity erodes exoplanet atmospheres[23-25] and affects potential habitability[26-27].

| Additional science questions addressed by pointed surveys: | | |
|---|---|---|
| How does plasma physics affect galaxy clusters? | X-ray spectral imaging of nearby clusters | [11,28] |
| How does feedback affect galaxy protoclusters? | X-ray imaging of z~3 Lyα emitters | [29] |
| What physics mediates supernova (SN) shocks? | Cadenced X-ray imaging of SN remnants | [30] |

***How do black holes merge?*** **TIME-DOMAIN SCIENCE**

LIGO has revealed a surprising variety of stellar mass black hole mergers[31], and LISA is poised to do the same for SMBHs[32,33]. Since SMBH mergers occur in dense and/or gas-rich nuclei, X-rays provide the crucial astrophysical context to LISA signals telling us that mergers have or are about to happen. X-rays directly trace the inner accretion disk where mergers and impacts occur, and where most of the accretion power is released. X-ray signals may include spectral signatures of mini-disks around binary SMBHs, variable X-ray signals from binary SMBHs due to Doppler boosting[34], flares as inspiraling stars or compact objects pass through the accretion disk (possibly already detected as quasi-periodic eruptions[35]), post-merger flares as the accretion disk reacts[36], and more[37]. Some transient X-ray signals, such as tidal disruption events (TDEs), may *anticipate* LISA detections, while other signals would be the subject of *searches* based on LISA signals. Either case requires cadenced time-domain surveys of wide fields or to repeatedly scan LISA error circles, with rapid reaction to LISA merger signals.

| Additional science questions addressed by time-domain capabilities: | | |
|---|---|---|
| How are the highest energy particles energized? | Fast X-ray followup of IceCube $\nu$ flares | [38] |
| What fraction of double white dwarfs explode? | Cadenced survey of Galactic Plane/Bulge | [30] |
| What are the progenitors of CC supernovae? | Real-time detection of SN shock breakout | [38] |

## 2. Science Table

Here we sketch “A” and “B” observing strategies for key questions to outline the instrument trade space. “A” is more direct, robust, and demands more from the instrument. “B” could convincingly answer them with an instrument that precludes “A”, but with greater science risk.

| | Required Measurement | Observing Program | $t_{exp}$ Instrument Dependence | Strict Requirements |
|---|---|---|---|---|
| *How did supermassive black holes form? (CoPAG and PhysPAG)* | | | | |
| A | AGN space density at 10 < z < 12 at 0.5-2 keV $L_X$ > $5 \times 10^{42}$ erg $s^{-1}$ | Observe $10^6$ $Mpc^3$ to 0.5-2 $F_X$ > $2 \times 10^{-18}$ erg $s^{-1}$ $cm^{-2}$ at 10 < z < 12 | $t_{exp}$ ~20 Ms × (Sky area/1 $deg^2$) × (FOV/0.1 $deg^2$)$^{-1}$ × ($A_{eff}$/1 $m^2$)$^{-1}$ × (PSF/1")$^2$ | -PSF ≤ 1.0" (sensitivity and source confusion) |
| B | Change in AGN space density from 8 < z < 10 in bins of Δz ≤ 1 and at 0.5-2 keV $L_X$ > $10^{43}$ erg $s^{-1}$ | Observe $10^8$ $Mpc^3$ to 0.5-2 $F_X$ > $10^{-17}$ erg $s^{-1}$ $cm^{-2}$ (3σ) at 8 < z < 10 | $t_{exp}$ ~45 Ms × (Sky Area/8 $deg^2$) × (FOV/0.1 $deg^2$)$^{-1}$ × ($A_{eff}$/1 $m^2$)$^{-1}$ × (PSF/1")$^2$ | -PSF ≤ 1.8" (sensitivity and source confusion) |
| *How does AGN feedback work in individual galaxies? (CoPAG)* | | | | |
| A | AGN-jet cavity power, $L_{cav}$, in galaxies to $L_{cav}$ > 25% $L_X$ within cooling radius | Measure PdV work to inflate cavity from n, T, $R_{cav}$ for 50 galaxies of 0.5-10 $L_*$. | $t_{exp}$ ~3 Ms × (N/50 gal) × ($A_{eff}$/1 $m^2$)$^{-1}$ | -PSF ≤ 1.5" (Error in $R_{cav}$, confusion) |
| B | | As above, for 30 early-type galaxies of 3-10 $L_*$. | $t_{exp}$ ~2 Ms × (N/30 gal) × ($A_{eff}$/1 $m^2$)$^{-1}$ | -PSF ≤ 3" (Error in $R_{cav}$, confusion) |
| *How and when was the Universe enriched with metals? (CoPAG)* | | | | |
| A | Galaxy cluster metallicity and scatter at 1 < z < 3 in bins of Δz ≤ 0.5 | Measure temperature and metallicity profiles from R = 0 to $R_{500}$ for 10 clusters/bin | $t_{exp}$ ~4 Ms × (Δz/0.5) × ($A_{eff}$/1 $m^2$)$^{-1}$ | -PSF ≤ 1.5" (AGN contamination) NXB ≤ 1.3x *Chandra* |
| B | | Measure the average metallicity for ≳30 Coma progenitors/bin | $t_{exp}$ ~8 Ms × (Δz/0.5) × ($A_{eff}$/1 $m^2$)$^{-1}$ [FOV > 0.7 $deg^2$ gives >1 cluster per FOV] | -PSF ≤ 3" (AGN contamination) NXB ≤ 1.3x *Chandra* |
| *How does high-energy stellar activity affect exoplanet atmospheres? (ExoPAG)* | | | | |
| A | Characterize stellar $L_X/L_{bol}$ and flares across age and spectral type. | Survey five Galactic Open Clusters to $F_X$>$10^{-17}$erg/s/$cm^2$ ($10^{27}$erg/s @ 1kpc) | $t_{exp}$ ~4 Ms × (FOV/0.1 $deg^2$)$^{-1}$ × ($A_{eff}$/1 $m^2$)$^{-1}$ [No efficiency at FOV>0.3 $deg^2$] | -PSF ≤ 1.8" (source confusion) |
| B | | Observe 50 GKM stars (0.1-3 Gyr) for ~3 days to measure quiescent $L_x$ and flare rates | $t_{exp}$ ~ 0 Ms × (N/50 stars) | -$A_{eff}$ > 2000 $cm^2$ (ability to measure quiescent $L_x$) |
| *What drives repeating nuclear X-ray transients? (PhysPAG)* | | | | |
| A | Determine the occurrence of QPEs at $L_X$ > $10^{42}$ erg $s^{-1}$ for 100 X-ray selected TDEs (200 $Gpc^{-3}$ $yr^{-1}$, peak $L_X$ ~ $10^{44}$ erg $s^{-1}$) | 2-yr monitoring of 0.5 $Gpc^3$ (14 day cadence) for X-ray selected TDEs (200 $Gpc^{-3}$ $yr^{-1}$, peak $L_x$ ~ $10^{44}$ erg $s^{-1}$) [at z=0.5, 280 $deg^2$ ~0.2 $Gpc^3$] | $t_{exp}$ ~1.7 ks × ($A_{eff}$/1 $m^2$)$^{-1}$ per FOV, per epoch at z ~ 0.5 $t_{exp}$ ~25 Ms × ($A_{eff}$/1 $m^2$)$^{-1}$ × (FOV/1.0 $deg^2$)$^{-1}$ × (Sky area/280 $deg^2$) × (Δt/2 weeks) × (t/2 yr) | -$A_{eff}$ > 3000 $cm^2$ (QPE duration) -FOV > 0.7 $deg^2$ (TDE rate) -PSF < 4" (confusion) |
| B | As above for 50 optically selected TDEs at z < 0.7 | Biweekly monitoring of TDEs over 2 years to 0.5-2 keV $F_X$ > $3 \times 10^{-15}$ erg $s^{-1}$ $cm^{-2}$ | $t_{exp}$ ~4 ks × ($A_{eff}$/1 $m^2$)$^{-1}$ per epoch $t_{exp}$ ~10 Ms × ($A_{eff}$/1 $m^2$)$^{-1}$ × (N/50 TDEs) × (Δt/2 weeks) × (t/2 yr) | -$A_{eff}$ > 3000 $cm^2$ (QPE duration) -PSF ≤ 4" (source confusion) |
| *What happens when two SMBHs merge? (PhysPAG)* | | | | |
| A | Search for X-rays at peak $L_X$ > $10^{44}$ erg $s^{-1}$ near the moment of merger at z=4 | Rapidly and repeatedly scan the LISA error circle (1-30 $deg^2$) for impending mergers | $t_{exp}$ ~0.3 Ms × (Sky area/10 $deg^2$) × ($N_{scans}$/3) × (FOV/0.3 $deg^2$)$^{-1}$ × ($A_{eff}$/1 $m^2$)$^{-1}$ + follow-up | -FOV > 0.3 $deg^2$, -$t_{ToO}$ < 10 min (transient duration) |
| B | As above, for peak $L_X$ > $10^{44}$ erg $s^{-1}$ at z=3 | Scan once and follow up the most promising signals | $t_{exp}$ ~0.1 Ms × (Sky area/10 $deg^2$) × (FOV/0.1 $deg^2$)$^{-1}$ × ($A_{eff}$/1 $m^2$)$^{-1}$ + follow-up | -FOV > 0.1 $deg^2$, -$t_{ToO}$ < 1 hr (transient duration) |

## 3. Instrument Description

The science priorities described above require an X-ray imaging instrument with 1”–2” PSF, large collecting area, sizable field of view, and moderate spectroscopic capabilities.

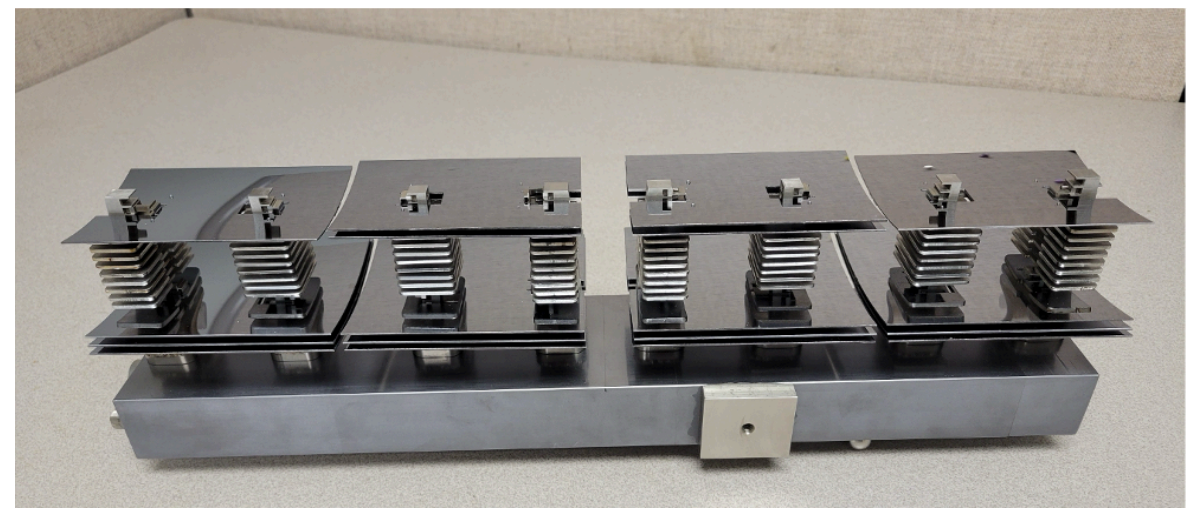

The baseline X-ray optic design uses the segmented monocrystalline silicon mirror technology developed by the Next Generation X-ray Optics (NGXO) team at NASA Goddard Space Flight Center[39]. The modular, hierarchical architecture of this technology enables the construction of large-aperture (>1 m), high-effective-area (up to 2 m²), high-angular-resolution (<2") mirror assemblies—a combination not currently achievable with replicated mirror technologies without employing multiple separate optics as was done for XMM-Newton. For a multiple-optic version, MSFC offers the replicated optic technology. Another option capable of similarly large-aperture mirrors is the silicon pore optics (SPO) being developed for ESA's NewAthena flagship mission[40]; however, within the next decade SPO is limited to an angular resolution of approximately 9". The Si mirror technology was significantly matured during the AXIS Phase A study, demonstrating high yields (>70%) of segments with ≤1" HPD[41]. The NGXO has also successfully compensated for stresses introduced by coating mirrors[42] and completed the first flight hardware realizations in the form of two engineering demonstration units (see figure). The primary remaining technical challenge involves mounting stress effects, which currently degrade performance from individual segment quality. The technology maturation pathway is strategically designed around a decision point in approximately 2030, with funding being pursued through the ASTRA initiative to demonstrate a TRL-5 module achieving ≤2" HPD performance. Critically, a comprehensive industrial feasibility study for mass production of segments and module assembly is required to establish production scalability and cost models; however, this essential study has not yet been budgeted or initiated, representing a gap that must be addressed for any future mission implementation.

The X-ray camera design is driven by the need for low-noise, fast operation. The AXIS High-Speed Camera[43] design meets many of the requirements for the mission concepts presented below, in particular the pixel size (0.55” at 9m focal length), field of view, and spectral performance. For large optics, detector frame rate becomes a major driver, as X-ray signals must be read out quickly enough that photons do not pile up. The CCDs designed and fabricated by MIT for AXIS, paired with readout ASICs developed by Stanford University, meet the speed requirements of AXIS; however, additional bright source and low surface brightness science can be realized by running much faster. Continued resources are required to advance the performance of this technology. Other technologies such as hybrid and monolithic silicon CMOS detectors[44] are viable for the range of concepts, given sufficient resources to reduce noise or increase broad-band X-ray performance. High-speed X-ray detectors require high-speed read-out, digitization, and event finding, and development of advanced flight electronics is also needed to realize this concept[45]. This includes advanced algorithms to reduce the effects of background[47-48], key to the low-surface-brightness science outlined above.

## 4. Mission Implementation

| | Bandpass (keV) | Field of View (arcmin) | Effective Area @ 1keV ($cm^2$) | Angular Resolution | | Response time | |
|---|---|---|---|---|---|---|---|
| | | | | On-axis (“ HPD) | FOV-ave (“ HPD) | On target following alert | TAM alerts to community |
| AXIS-Classic | 0.3-10 | 27x27 | 5000 | 1.5 | 1.8 | <1 hr | <10 min |
| AXIS-Fast | 0.2-6 | 60x60 | 6000 | 2.0 | 3.5 | <10 min | <3 min |
| AXIS-Deep | 0.3-10 | 20x20 | 5000 | 1.0 | 1.3 | <1 hr | <10 min |

Table 2 : Baseline requirements of the three strawman concepts.

Our three concepts can be assessed against scientific priorities and technical readiness in the early 2030s for launch in the late-2030s. All concepts combine high-angular resolution X-ray spectral imaging with a TDAMM-responsive platform deployed at Sun-Earth L2 (85% of the sky observable). They differ in emphasis between deep imaging and TDAMM, distinguished by their baseline angular resolution, field of view, and response/alert time (Table 2).

**AXIS-Classic** is very similar to the AXIS Phase A baseline as submitted to NASA's Astrophysics Probe Explorer (APEX) call. It executes a balanced program of surveys, targeted observations and targets of opportunity (ToOs) discussed in Section 1. Assuming that 1.5” (half-power diameter; HPD) X-ray optics have been matured by 2030, the AXIS Phase A study shows that the AXIS X-ray Mirror Assembly can be built and the mission deployed within six years at a cost of $1.2B plus launch costs (FY30; corresponding to $1.0B FY23). *AXIS-Classic would be a natural component of a multi-wavelength Flaglet Fleet.*

**AXIS-Fast** is optimized for time-domain science. A focal length of 4.5m (c.f. 9m for AXIS-Classic/AXIS-Deep) enables a large (1$deg^2$) field-of-view in the 0.2-6keV band. The angular resolution needs are relaxed to 2.0” (on-axis) / 3.5” (FOV-average). The ground-segment, comms, and spacecraft agility are optimized for fast response to community alerts (including GW and neutrino triggers). Operations emphasize cadenced surveys of Roman extragalactic/ Galactic fields and ToOs, including localization of LISA-detected merging SMBHs. *AXIS-Fast would be a natural component of a multi-wavelength TDAMM Fleet.*

**AXIS-Deep** is optimized for point source sensitivity and, at 1.0” (HPD; on-axis), is the most demanding of the X-ray optics and detector technology. Operations emphasize deep extragalactic surveys focused on early SMBH growth; targeted observations of nearby and distant galaxies, galaxy groups, and clusters to probe feedback and enrichment; surveys of open clusters for the host-exoplanet connection; and ToOs on well-localized transients.

**Partnerships and synergy with Industry** : During the Phase A study, the AXIS team identified several industry partners who had developed many of the technologies needed for polishing, coating and mounting the silicon mirrors. The selected partner (Attalon, formerly known as Tinsley) additionally developed costed plans for the manufacturing of the mirror segments and modules. It is clear that industry will be an essential partner in future large-throughput X-ray observatories, and has the ability and willingness to engage with NASA and mission teams.